# Highly-Reverberant Real Environment database: HRRE

*Juan Pablo Escudero*[1], *Victor Poblete*[2], *José Novoa*[1], *Jorge Wuth*[1], *Josué Fredes*[1], *Rodrigo Mahu*[1], *Richard Stern*[3] and *Néstor Becerra Yoma*[1]

[1] Speech Processing and Transmission Laboratory, Electrical Engineering Department, University of Chile, Santiago, Chile.
[2] Institute of Acoustics, Universidad Austral de Chile, Valdivia, Chile.
[3] Department of Electrical and Computer Engineering and Language Technologies Institute, Carnegie Mellon University, Pittsburgh, USA.

nbecerra@ing.uchile.cl

**Abstract**

Speech recognition in highly-reverberant real environments remains a major challenge. An evaluation dataset for this task is needed. This report describes the generation of the Highly-Reverberant Real Environment database (HRRE). This database contains 13.4 hours of data recorded in real reverberant environments and consists of 20 different testing conditions which consider a wide range of reverberation times and speaker-to-microphone distances. These evaluation sets were generated by re-recording the clean test set of the Aurora-4 database which corresponds to five loudspeaker-microphone distances in four reverberant conditions.

## 1. Introduction

Several challenges or databases have been generated to address the problem of reverberated speech in ASR CHiME-2 [17] includes a medium vocabulary task where clean utterances were convolved with binaural room impulse responses that were estimated for 121 different positions covering a horizontal square grid of 20 cm side centered at 2 m from the microphones. CHiME-3 [2] is composed of data collected in a real-world scenario where a person talks to a tablet device, which has been fitted with 6-channel microphone array. Several real environments such as café, street junction, public transport and pedestrian area were employed. CHiME-4 [3] is the same data as CHiME-3 but increases the level of difficulty by constraining the number of microphones for testing. The CHiME-5 challenge targets the problem of distant microphone speech recognition in everyday home environments. Speech material has been collected from twenty real dinner parties that have taken place in real homes using multiple 4-channel microphone arrays. The REVERB challenge [15] includes simulated (SimData) and real data (RealData). SimData emulates six different reverberation conditions: three rooms with different volumes (small, medium and large), two different distances between a speaker and a microphone array (near=50 cm and far=200 cm). The reverberation time (RT) corresponding to the small, medium, and large-size rooms are about 0.25 s, 0.5 s and 0.7 s, respectively. In contrast, RealData contains a set of real recordings in a reverberant meeting room that are different from those used for SimData. The reverberant meeting room has a RT equal to 0.7 s. RealData contains two reverberation conditions: one room, two types of distances between the speaker and the microphone array (near, 100 cm, and far, 250 cm). In the case of the ASpIRE challenge [18] the data was collected using a set of different microphones placed in a wide range of locations in seven different rooms (some classrooms and some office space) with several different shapes, sizes, surface properties, and noise sources. Speakers were recorded from different positions in each room.

Observe that all the reverberated databases that have been employed so far attempt to use real environments and, in most cases, also include additive noise. Surprisingly, the response of the ASR technologies to RT and speaker-microphone distance has not been addressed methodologically and independently of the additive noise yet. This is partly because there has not been a suitable database for this purpose. HRRE is a response to this need by providing speech datasets recorded in a controlled reverberant environment with several values for the speaker-microphone distance. By doing so, we are covering a wide range of potential applications that span all over from human-robot applications, meeting rooms, smart houses to close-talk microphone scenarios.

## 2. Database Recording

To generate the data for the test set, we re-recorded the original clean test data from the Aurora-4 database (*i.e.* 330 utterances recorded with the Sennheiser microphone) in a reverberation chamber considering different speaker-microphone distances and RTs and following the procedures specified by the ISO 354:2003 Standard [1]. The reverberation chamber has an internal surface area of 100 m$^2$, a volume of 63 m$^3$ and an RT$_{mid}$ equal to three seconds. Four reverberant conditions were generated by adding sound-absorbing materials in the reflecting surfaces of the chamber (see Fig. 1). The corresponding RT$_{mid}$ were computed according to the specifications of the ISO 3382-1:2009 Standard [2]. The resultant RT$_{mid}$ values were 0.47 s, 0.84 s, 1.27 s and 1.77 s, which are the RTs suggested in the literature for meeting rooms and multipurpose halls [3] [4].

The loudspeaker-microphone distances were selected as follows. The longest distance was set to 2.56 m. Then, the distance was reduced four times by factors of two ultimately reaching 0.16 m, as can be seen in the Fig. 2. Following this procedure, the selected distances were: 0.16 m, 0.32 m, 0.64 m, 1.28 m and 2.56 m. The recordings were obtained using an HP Probook laptop, a Samsom Servo 260 power amplifier, a Bose V 201a loudspeaker, an Earthworks M30 measurement

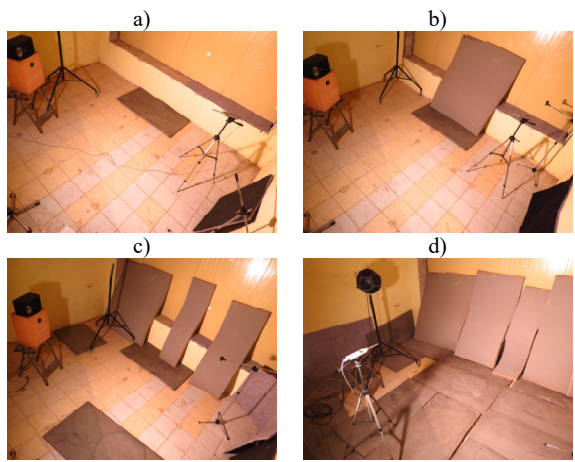
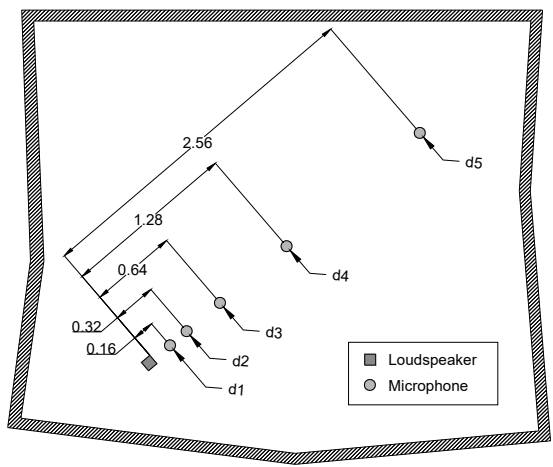

Figure 1: *Experimental setup in our reverberation chamber a) RT= RT=1.77 s, b) RT=1.27 s, c) RT=0.84 s and d) 0.47 s.*

Figure 2: *Recording scheme of distances used in the reverberation chamber. The selected loudspeaker-microphone distances were: $d_1$=0.16 m, $d_2$=0.32 m, $d_3$=0.64 m, $d_4$=1.28 m and $d_5$=2.56 m.*

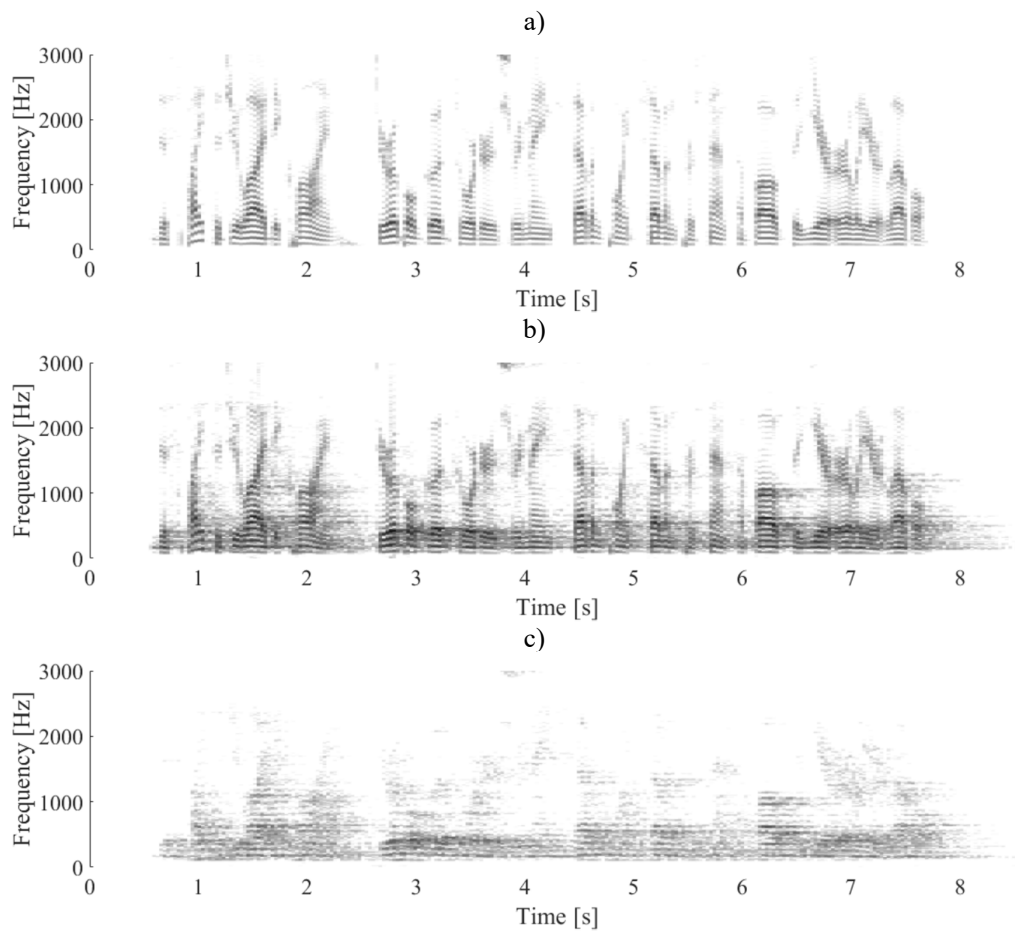

Figure 3: *Spectrograms for a sample utterance, a) the clean utterance, the recorded utterance with RT=1.77 s at b) d=0.16 m and c) d=2.56 m.*

microphone and Focusrite Scarlett 2i2 audio interface.

The loudspeaker power level was previously set in an anechoic chamber. The level was adjusted to reach an equivalent sound pressure level (Leq) of 60 dBA one meter away from the loudspeaker over one minute, according to the IEC 60268:2003 Standard [5]. For this measurement, the loudspeaker reproduced ten utterances from different speakers. The measurement was performed using the Cesva SC 310 sound level meter.

The recording level in the reverberation chamber was set using the same utterances that had been used for setting the loudspeaker power level. The recording level was set to exceed the background noise and to be below the audio interface clipping level. This procedure was performed at the 0.32 m and 2.56 m loudspeaker-microphone distances.

The background noise was kept under control and measured before recording the test database at each condition. The maximum background noise Leq over one minute was equal to 37 dBA.

Figure 3b and 3c show spectrograms of a sample utterance for different distances in the most reverberant environment, RT=1.77 s. These spectrograms are contrasted with the spectrogram of the clean utterance shown in Fig. 3a. The delay produced by the experimental setup was compensated using cross correlation to align the beginnings of the re-recorded utterances with the original utterances.

In the setup described, the clean test utterances of the Aurora-4 database were re-recorded in 20 different testing conditions, which correspond to each selected reverberant environment and loudspeaker-microphone distance pair. As a result, 13.4 hours of data recorded in highly reverberant real environments were obtained. Instructions for requesting the HRRE data are available to the general public at http://www.lptv.cl/en/hrre/.

## 3. Acknowledgements

The research reported here was funded by Grants Conicyt-Fondecyt 1151306 and ONRG N°62909-17-1-2002. José Novoa was supported by Grant CONICYT-PCHA/Doctorado Nacional/2014-21140711. The authors would also like to thank the Institute of Acoustics, Universidad Austral de Chile, for providing the reverberation chamber and the equipment required to record this database.